\documentclass[twocolumn,aps,pra,showpacs,floatfix,longbibliography]{revtex4-1}

\usepackage{latexsym}
\usepackage{amsmath}
\usepackage{amssymb}
\usepackage{latexsym}
\usepackage{color}
\usepackage{graphicx}
\usepackage{cancel}
\usepackage{bbm}
\usepackage{bm}
\usepackage{maybemath}

\newcommand{\ii}{{\mathrm i}}
\newcommand{\ee}{{\mathrm e}}

\newcommand{\calM}{\mathcal{M}}

\newcommand{\calP}{\mathcal{P}}
\newcommand{\calQ}{\mathcal{Q}}

\newcommand{\calV}{\mathcal{V}}

\begin{document}

\title{Non--Resonant Two--Photon Transitions in Length and Velocity Gauges}

\author{U. D. Jentschura}
\affiliation{Department of Physics,
Missouri University of Science and Technology,
Rolla, Missouri 65409, USA}

\begin{abstract}
We reexamine the invariance of 
two-photon transition matrix elements and 
corresponding two-photon Rabi frequencies under 
the ``gauge'' transformation from the 
length to the velocity gauge. 
It is shown that gauge invariance, in the most 
general sense,  only holds at exact resonance,
for both one-color as well as two-color absorption.
The arguments leading to this conclusion are 
supported by analytic calculations which 
express the matrix elements in terms of hypergeometric
functions, and ramified by
a ``master identity'' which is fulfilled by 
off-diagonal matrix elements of the Schr\"{o}dinger 
propagator under a the transformation from the 
velocity to the length gauge.
The study of the gauge dependence of atomic processes
highlights subtle connections between the
concept of asymptotic states, the gauge
transformation of the wave function,
and infinitesimal damping
parameters for perturbations and interaction
Hamiltonians that switch off the terms in the
infinite past and future [of the form $\exp(- \epsilon | t |)$].
We include a pertinent discussion.
\end{abstract}


\maketitle

%
%
\section{Introduction}
\label{sec1}

A priori, the description of any dynamic process in atomic physics 
should be gauge invariant. Yet, there is a caveat. 
Namely, one usually calculates
atomic transitions using wave functions obtained from the
solution of the unperturbed Schr\"{o}dinger equation,
and ignores both the gauge transformation of the wave function
as well as the fact that the physical
interpretation of the wave function changes
under a gauge transformation
from the velocity to the length gauge.
Lamb~\cite{La1952} has shown that if one
insists on using ordinary Schr\"{o}dinger wave functions
in off-resonant one-photon transition matrix elements,
then the length gauge form has to be used for the laser field.
Here, we refer to off-resonant transitions as those where the 
frequency of the incident radiation 
is not exactly equal to the resonance frequency of the atom;
generalized to two-photon transitions, this implies 
that the sum frequency of the two photons 
does not exact match the energy (frequency) difference
of the ground and excited level.

For off-resonant two-photon transitions, one might 
think that the transition matrix element could be 
equal in the length and velocity gauges if all possible intermediate,
virtual states are included in the calculation.
Here, inspired by Refs.~\cite{BaFoQu1977,Ko1978prl,ScBeBeSc1984},
we aim to reinvestigate the status of the 
gauge invariance of two-photon transition matrix elements
(``length'' versus ``velocity'' gauges).
Note that invariance under the change from the 
``length'' to the ``velocity'' gauges implies that 
the gauge transformation of the
wave function is ignored (we also refer to the invariance
of matrix elements under the neglect of the
wave function transformation and under the
neglect of any necessary reinterpretation
of physical operators as the ``extended gauge
invariance'').
In Refs.~\cite{PoTh1975,PoTh1978},
the two-photon transition matrix element
has been examined without any additional condition 
enforced upon the laser frequency;
the laser might be off-resonant or on resonance.
As is explained below, the arguments given in 
Refs.~\cite{PoTh1975,PoTh1978} appear to be applicable 
only at exact resonance.
The role of the intermediate,
virtual quantum states in the gauge invariance
will be analyzed here in terms of 
general identities, applicable to various 
physical processes.
Second, from a conceptual point of 
view, the role of the gauge transformation of the 
wave function and the concomitant change in its
interpretation appears to profit 
from further explanatory remarks beyond the problem at hand.

The subject matter of this article is rather basic 
quantum mechanics, supplemented 
with explicit analytic results for the 
length and velocity forms of the $1S$--$2S$ two-photon
transition matrix element in hydrogen.
We start by investigating 
gauge transformations and physical interpretations 
of operators in Sec.~\ref{sec2},
before reexamining the ``gauge invariance''
of the ac Stark shift under a
change of the interaction Hamiltonian from the 
length to the velocity form (Sec.~\ref{sec3}).
Here, ``gauge invariance'' has to taken with a grain
of salt; the invariance of the theoretical expressions for the 
ac Stark shift holds even if the mandatory gauge transformation of the 
wave function is neglected, a fact on which we comment in 
Sec.~\ref{sec4}.
In Sec.~\ref{sec3}, we shall examine, based on explicit 
analytic and numerical calculations, the 
behavior of a typical two-photon 
transition matrix element 
(namely, of the $1S$--$2S$ two-photon transition in hydrogen)
off resonance.
We aim to show that the inclusion of the intermediate states
does not solve the problem of gauge invariance,
but the gauge dependence is due to a change in the physical interpretation of the
wave function under the presence of a nonvanishing
vector potential. The physically correct result for the
transition rate off resonance is obtained in the length gauge.

%
%
\section{Gauge Transformation and Physical Interpretations}
\label{sec2}

In order to illustrate that 
gauge transformations can change the physical interpretation of 
operators, let us start from a trivial example.
We consider a wave function $\psi(\vec r) = 1/\sqrt{V}$
where $V$ is the normalization volume;
it describes a particle at rest. 
A unitary ``gauge'' transformation of the form
\begin{equation}
\psi( \vec r) \to \exp\left(\frac{\ii}{\hbar} \vec p_0 \cdot \vec r\right) \, 
\psi(\vec r)
\end{equation}
is applied. The momentum operator in the 
free Hamiltonian $H = \vec p^{\,2}/(2 m)$,
with $\vec p = -\ii \vec\nabla$, transforms as 
\begin{equation}
\vec p \to \ee^{\frac{\ii}{\hbar} \vec p_0 \cdot \vec r} \;
\vec p \; \ee^{-\frac{\ii}{\hbar} \vec p_0 \cdot \vec r}
= \vec p - \vec p_0 \,.
\end{equation}
The gauge-transformed Hamiltonian thus 
reads as $H' = (\vec p - \vec p_0)^2/(2 m)$,
and the interpretation of the 
momentum operator $\vec p = -\ii \hbar \vec \nabla$
has changed: Namely, the 
kinetic momentum operator no longer is 
$\vec p$, but $\vec p - \vec p_0$.
Indeed, 
$\vec p_{\rm kin} = \vec p - \vec p_0$ is the conjugate variable 
of the position operator $\vec r$
(see also the Appendix of Ref.~\cite{JeNo2013pra}).

A similar situation is encountered in 
electrodynamics~\cite{Re2013}. The time-dependent wave 
function receives a unitary gauge transform,
\begin{equation}
\psi(t, \vec r) \to \ee^{\frac{\ii}{\hbar} e \Lambda(t, \vec r)} \, \psi(t, \vec r) \,.
\end{equation}
The momentum operator transforms as 
\begin{align}
\vec p \to & \; 
\ee^{\frac{\ii}{\hbar} e \Lambda(t, \vec r)} \,
\vec p \, \ee^{-\frac{\ii}{\hbar} e \Lambda(t, \vec r)} \,
\to \vec p - e \vec \nabla \Lambda(t, \vec r) \,,
\end{align}
where $e$ is the electron charge.
The following Hamiltonian describes the 
atom-electromagnetic field dynamical system
consisting of an electron coupled to a 
vector potential $\vec A$,
in the binding (Coulomb) potential $\Phi$,
\begin{equation}
\label{dyn}
H = \frac{(\vec p - e \, \vec A)^2}{2 m} + e \Phi \,.
\end{equation}
The product $e \Phi$ of electron charge and
binding scalar potential is often denoted as $V$,
because it acts as a potential term in the Hamiltonian.
Typicall, the vector potential $\vec A$ describes a laser.
The unitary gauge transformation, applied to $H$,
leads to  the transformed Hamiltonian $\widetilde H$,
\begin{equation}
\widetilde H = 
\ee^{\frac{\ii}{\hbar} e \Lambda(t, \vec r)} \,
H \, \ee^{-\frac{\ii}{\hbar} e \Lambda(t, \vec r)}
= \frac{(\vec p - e \, \vec A')^2}{2 m} + e \Phi \,,
\end{equation}
where $\vec A' = \vec A + \vec\nabla \Lambda$ 
is the gauge-transformed vector potential.
Conversely, for $\vec A = \vec 0$, and $\vec A' = \vec\nabla \Lambda \neq \vec 0$,
one asserts that the interpretation of the 
momentum operator $\vec p = -\ii\hbar \vec\nabla$ changes;
it no longer describes the kinetic momentum.
The place of the latter is taken by the conjugate 
variable of position, namely, $\vec p_{\rm kin} = \vec p - e \, \vec A'$.

However, under the gauge transformation, the 
physical interpretation of the Hamiltonian 
also changes. A priori, the Hamiltonian $H$
is equal to the the time derivative operator $\ii \hbar \partial_t$.
After the gauge transformation, it is 
equal to a unitarily transformed time 
derivative operator,
\begin{equation}
\ii \hbar \partial_t 
\to \ee^{\frac{\ii}{\hbar} e \Lambda(t, \vec r)} \,
(\ii \hbar \partial_t) \, \ee^{-\frac{\ii}{\hbar} e \Lambda(t, \vec r)}
\to \ii \hbar \partial_t + e \partial_t \Lambda(t, \vec r) \,.
\end{equation}
The new time derivative operator is thus 
obtained by setting equal the 
unitarily transformed Hamiltonian and the 
unitarily transformed time derivative 
operator, and reads as
\begin{equation}
H' = \widetilde H - e \partial_t \Lambda = 
\frac{(\vec p - \vec A')^2}{2 m} + e \, \Phi' \,,
\end{equation}
where $\Phi' = \Phi - \partial_t \Lambda$ is the 
gauge-transformed scalar potential.

From the above derivation, which in principle 
recalls well-known facts, it is 
immediately obvious that the gauge-transformed Hamiltonian 
$H'$ cannot be obtained from $H$ by a unitary transformation.
Conversely, the unitarily transformed $\widetilde H$ cannot be
interpreted any more as the time derivative operator
when acting on the gauge-transformed wave function.

In order to fix ideas, it is useful to examine the 
velocity-gauge one-photon transition matrix element
$\calM_v$,
\begin{align}
\label{fix1}
\calM_v =& \; \left< \phi_f \left| 
-e \, \frac{\vec A \cdot \vec p}{m} 
+\frac{e^2 \vec A^{\,2}}{2 m} 
\right| \phi_i \right>
\nonumber\\[0.133ex]
\to & \; \left< \phi_f \left|
-e \, \frac{\vec A \cdot \vec p}{m}
\right| \phi_i \right>
= \left< \phi_f \left|
-e \, \frac{\ii}{\hbar} \, \vec A \cdot [ H, \vec r] 
\right| \phi_i \right>
\nonumber\\[0.133ex]
= & \; 
-e \, \frac{\ii}{\hbar} (E_f - E_i) \,
\left< \phi_f \left| \vec A \cdot \vec r \right| \phi_i \right>
\end{align}
of an initial atomic state $| \phi_i \rangle$ and a 
final state $|\phi_f\rangle$.
[We have made the dipole approximation $\vec A = \vec A(t)$ 
and assumed that the seagull term 
$e^2 \vec A^{\,2}/(2 m)$ does not contribute 
because of the angular symmetry of the initial
and final states involved in the dipole transition.]
In view of the identity $\vec E = -\partial_t \vec A \to
\ii \omega \, \vec A$, the length-gauge one-photon matrix 
elements $\calM_\ell$ is related to its velocity-gauge 
counterpart $\calM_v$ as follows,
\begin{align}
\label{fix2}
\calM_\ell 
=& \; \left< \phi_f \left| 
-e \, \vec E \cdot \vec r \right| \phi_i \right>
\nonumber\\[0.133ex]
\to & \; 
- \ii \omega \, \frac{e}{m} \,
\left< \phi_f \left| 
\vec A \cdot \vec r \right| \phi_i \right> 
= \frac{\hbar \omega}{E_f - E_i} \, \calM_v \,.
\end{align}
The velocity-gauge expression $\calM_v$ differs from its 
length-gauge counterpart $\calM_\ell$
by an additional factor $(E_f - E_i)/\omega$.
Hence, in a remark on p.~268 of Ref.~\cite{La1952},
Lamb observed that the physical interpretation
of the wave function is preserved only in the 
length gauge, 
and ``no additional factor $(E_f - E_i)/\omega$
actually occurs''.
Indeed, the physical interpretation of the momentum operator 
in the Schr\"{o}dinger--Coulomb Hamiltonian 
is preserved only in the length gauge
after a laser field is switched on.
In other words, the problems off resonance
with the velocity gauge result from the fact
that one uses a wave function,
which is an eigenstate of a Hamiltonian that involves
the momentum operator $\vec p$, 
and formulates the interaction Hamiltonian 
$-e \vec A \cdot \vec p/m + e^2 \vec A^{\,2}/(2 m)$
with an expression that also involves the 
momentum operator $\vec p$, but in a 
situation where $\vec p$ loses the original
physical interpretation that it had in the unperturbed 
Schr\"{o}dinger--Coulomb Hamiltonian.
Alternatively, one can also argue that the electric field
used in the length-gauge interaction is gauge invariant,
while the vector potential in the velocity-gauge
term is not~\cite{Ko1978prl,ScBeBeSc1984,JeEvHaKe2003,%
JeKe2004aop,EvJeKe2004,JeEvKe2005}.

Within the dipole approximation, the atomic Hamiltonian
reads as follows
[we denote the laser field by $E(t)$ and the 
binding Coulomb potential by $V$]
\begin{equation}
H_{A\ell} = \frac{\vec p^{\,2}}{2 m} + V - e \, \vec E(t) \cdot \vec r \,.
\end{equation}
Under a gauge transformation
$\Lambda(t, \vec r) = \vec A(t) \cdot \vec r$,
this Hamiltonian is transformed to 
\begin{equation}
H_{Av} = \frac{[\vec p - e \, \vec A(t)]^2}{2 m} + V \,.
\end{equation}
As we have seen, the two Hamiltonians $H_{A\ell}$ and 
$H_{Av}$ are not related by a unitary transformation,
and furthermore, matrix elements of the 
interaction Hamiltonians 
\begin{equation}
H_\ell = - e \, \vec E(t) \cdot \vec r
\end{equation}
and 
\begin{equation}
H_v = - e \, \frac{\vec A(t) \cdot \vec p}{m} + 
\frac{e^2 \vec A(t)^{\,2}}{2 m}
\end{equation}
differ off resonance.
Despite this fact, a number of processes 
such as the ac Stark shift are 
``gauge-invariant'' (in an extended sense) 
under a replacement of the 
interaction $H_\ell$ by $H_v$, 
even without any gauge transformation 
of the wave function.

Before we discuss the reasons why 
``extended gauge invariance'' holds or fails
for a given physical problem, we shall
reexamine the extended gauge invariance of the 
ac Stark shift, and the failure of the 
extended gauge invariance in the case of the 
two-photon transition matrix element.

%
%
\section{Gauge Invariance of the ac Stark Shift}
\label{sec3}

The derivation of the ac Stark shift 
is easiest in a second-quantized formalism,
where the $z$-polarized laser field
in the dipole approximation is modeled by a field operator,
resulting in a length-gauge interaction
\begin{subequations}
\label{HLI}
\begin{align}
\vec E_L = & \; \hat e_z \, \sqrt{\frac{\hbar \omega}{2 \epsilon_0 \calV_L}} \,
\left( a_L + a_L^+ \right)
= \hat e_z \, E_L \,,
\\[0.133ex]
H_\ell =& \; -e \, z \, E_L  \,.
\end{align}
\end{subequations}
Here, the $a_L$ and $a^+_L$ are the annihilation and creation
operators for laser photons.
The velocity-gauge interaction is given as 
\begin{subequations}
\begin{align}
\vec A_L =& \; \hat e_z \, \sqrt{\frac{\hbar}{2 \omega \epsilon_0 \calV_L}} \,
\left( \ii a_L - \ii a_L^+ \right)
= \hat e_z \, A_L \,,
\\[0.133ex]
H^{(1)}_v =& \; -\frac{e \, A_L \, p_z}{m}  \,,
\qquad
H^{(2)}_v = \frac{e^2 \, A_L^2}{m}  \,.
\end{align}
\end{subequations}
The unperturbed Hamiltonian is the sum of the atomic Hamiltonian
$H_A$ and the Hamiltonian $H_{EM}$ which describes the 
electromagnetic field (laser mode $L$ and 
modes $\vec k \, \lambda$ other than the laser field),
\begin{subequations}
\begin{align}
H_0 =& \; H_A + H_{EM} \,,
\\[0.133ex]
H_A =& \;
\sum_m E_m \, | \phi_m \rangle \, \langle \phi_m | \,,
\\[0.133ex]
H_{EM} =& \;
\sum_{\vec k \, \lambda \neq L} \hbar \omega_{\vec k \, \lambda}
a_{\vec k \, \lambda}^+ \; a_{\vec k \, \lambda} +
\hbar \omega \; a^+_L \; a_L \,.
\end{align}
\end{subequations}
The unperturbed state $| \phi_0 \rangle = | \phi, \, n_L\rangle$ 
with the atom in state $|\phi \rangle$ and $n_L$ laser photons
fulfills the relationships
\begin{subequations}
\begin{align}
| \phi_0 \rangle =& \; | \phi, \, n_L\rangle  \,,
\qquad
H_A \, | \phi \rangle = E \, |\phi \rangle \,,
\\[0.133ex]
H_0 \, | \phi_0 \rangle =& \; E_0  \, | \phi_0 \rangle \,,
\qquad
E_0 = E + n_L \hbar \omega \,.
\end{align}
\end{subequations}
The reduced Green function for the combined 
system of atom and radiation field is given 
by 
\begin{equation}
G'(\omega) = 
\left( \frac{1}{H_0 - E_0} \right)' \,,
\end{equation}
where $H_0$ contains both atomic as well as 
laser-field terms and the prime on the Green function
denotes the omission of the reference state $| \phi_0 \rangle$
of the combined atom$+$field system from the sum 
over intermediate state.

In the length gauge, the second-order ac Stark shift
can be expressed as
\begin{align}
\label{Eell}
\Delta E = & \;
- \left< \phi_0 \left| H_L G'(\omega) \, H_L \right| \phi_0 \right>
= - \frac{I_L}{2 \epsilon_0} \, P(\phi) \,.
\end{align}
The laser-field intensity is
\begin{equation}
\label{IL}
I_L = \frac{n_L \hbar \omega \, c}{\calV_L} \,,
\end{equation}
and the polarizability is given as
\begin{equation}
\alpha(\omega) =
\frac{e^2}{3} \sum_{\pm}
\left< \phi \left| \vec r \, \frac{1}{H_A - E \pm \hbar \omega} \, \vec r
\right| \phi \right> \,.
\end{equation}
(We here assume a radially symmetric reference 
state~$| \phi \rangle$.)
In the language of second quantization~\cite{CrTh1984},
the two terms with an opposite sign of $\hbar \omega$ in 
the denominator 
are generated by paired photon annihilation and photon 
creation operators from Eq.~\eqref{HLI}.

The extended gauge invariance of the ac Stark shift
in atomic hydrogen relies on the fact that $\Delta E$ given in 
Eq.~\eqref{Eell} can alternatively be expressed as
\begin{equation}
\label{Ev}
\Delta E = 
 - \left< \phi_0 \left| H^{(1)}_v G'(\omega) \, 
H^{(1)}_v \right| \phi_0 \right> +
\left< \phi_0 \left| H^{(2)}_v \right| \phi_0 \right> \,.
\end{equation}

After treating the photons, the extended gauge 
invariance is easily shown to be equivalent to the 
identity 
\begin{multline}
\label{pol2}
\frac{1}{m} \sum_{\pm} \left< \phi \left| \vec p
\frac{1}{H_A - E \pm \omega}
\vec p \right| \phi \right> - 3 \, \left< \phi | \phi \right>
\\[2ex]
= \omega^2 \, 
\sum_\pm \left< \phi \left| \vec r \,
\frac{m}{H_A - E \pm \omega} \, \vec r \right| \phi \right> \,,
\end{multline}
where $\left< \phi | \phi \right> = 1$ is the normalization integral;
the corresponding term originates from the seagull 
Hamiltonian $e^2 A_L^2/(2 m)$.

In order to show this identity, we generalize the 
problem somewhat and write the following two matrix elements,
\begin{subequations}
\begin{align}
P(\omega) = & \; \frac{\hbar^2}{m^2} \left< \phi_f \left| \vec p  \, 
\frac{1}{H_A - E_n - \omega} \,
\vec p \right| \phi_i \right> \,,
\\[0.1133ex]
Q(\omega) = & \;
\left< \phi_f \left| \vec r \, 
\frac{1}{H_A - E_n - \omega} \, \vec r \right| \phi_i \right> \,,
\end{align}
\end{subequations}
for two (not necessarily equal) atomic states $|\phi_i \rangle$ and 
$|\phi_f \rangle$.
Repeated application of the commutator relations
\begin{equation}
\label{commutator}
\frac{\vec{p}}{m} = \frac{\ii}{\hbar} \, \, [H_A, \vec{r}] \,,
\qquad \qquad
H_A = \frac{\vec{p}^{\,2}}{2 m} - \frac{e^2}{4 \pi \epsilon_0 r}\,,
\end{equation}
results in the equality
\begin{multline}
\frac{\hbar^2}{m^2} \left< \phi_f \left| \vec p  \,
\frac{1}{H_A - E_i - \omega} \,
\vec p \right| \phi_i \right> 
\\[0.1133ex]
= (E_f - E_i - \hbar \omega) (-\hbar \omega)
\left< \phi_f \left| \vec r \, 
\frac{1}{H_A - E_i - \omega} \, \vec r \right| \phi_i \right>
\\[0.1133ex]
+ (\hbar \omega - E_f) \, \left< \phi_f \left| \vec r^{\,2} \right| \phi_i \right> 
+ \left< \phi_f \left| \vec r \, H_A \, \vec r \right| \phi_i \right> \,.
\end{multline}
One rewrites this expression using the 
operator identity
\begin{equation}
\vec r \, H_A \, \vec r =
\frac12 \, \left( 
[\vec r, \, [H_A, \, \vec r ]] + 
\vec r^{\,2} \, H_A +
H_A \, \vec r^{\,2} \right) \,,
\end{equation}
applies the Hamiltonian $H_A$ on either side 
to an eigenstate, and concludes that 
the following ``master identity'' holds,
\begin{align}
\label{genid}
P(\omega) & \; = (E_f - E_i - \hbar\omega) (-\hbar\omega) \, Q(\omega)
\nonumber\\[0.1133ex]
& \; + \left(\hbar\omega - \tfrac12 \, ( E_f - E_i) \right) \, 
\left< \phi_f \left| \vec r^{\,2} \right| \phi_i \right> 
\nonumber\\[0.1133ex]
& \; + \frac{3 \hbar^2}{2 m} \left< \phi_f | \phi_i \right> \,.
\end{align}
For the ac Stark shift, one sets $E_f = E_i = E$, 
$|\phi_f \rangle = |\phi_i \rangle = |\phi \rangle$, 
and adds two terms with $\pm \omega$.
One can thus easily show that Eq.~\eqref{pol2} 
follows from Eq.~\eqref{genid},
demonstrating the ``extended gauge invariance'' of the ac
Stark shift.
Recently, an analogous derivation has been shown to lead to the 
``extended gauge invariance'' of the one-loop correction to the 
imaginary part of the polarizability~\cite{JePa2015epjd1}.

%
%
\section{Two--Photon Matrix Element in Length and Velocity Gauge}
\label{sec4}

Here, the situation is different from the ac Stark shift;
the initial state consists of a combined 
atom$+$field state where the atom is in the 
ground state and $n_L + 2$ photons are in the laser
mode,
\begin{equation}
| \phi_1 \rangle = | \phi_i, n_L + 2 \rangle \,.
\end{equation}
The final state has the atom in state $| \phi_f\rangle$
and two photons less in the laser mode,
\begin{equation}
| \phi_1 \rangle = | \phi_f, n_L \rangle \,.
\end{equation}
The matrix element for the transition is
\begin{align}
\calM =& \; 
\left< \phi_2 \left| H_\ell \, G'(\omega) \, H_\ell \right| \phi_1 \right>
\nonumber\\[2ex]
\mathop{=}^{\mbox{?}} & \;
\left< \phi_2 \left| H^{(1)}_v \, G'(\omega) \, 
H^{(1)}_v \right| \phi_1 \right>
+ \left< \phi_2 \left| H^{(2)}_v \right| \phi_1 \right> \,.
\end{align}
Here, in contrast to the ac Stark shift,
only one term contributes in the electric field,
namely, the one with the 
annihilation operators.
Furthermore, because $\left< \phi_f | \phi_i \right> = 0$
(the two states are manifestly different),
the seagull term makes no contribution.
After treating the photon degrees of freedom,
the equality of the length and velocity gauge 
expressions for the two-photon matrix element
is easily shown to be equivalent to the relation
\begin{multline}
\label{twogamma1}
\frac{1}{m^2} \left< \phi_f \left| \vec p \,
\frac{1}{H_A - E - \omega}
\, \vec p \right| \phi_i \right> 
\\[2ex]
\mathop{=}^{\mbox{?}} \pm \omega^2 \,
\left< \phi_f \left| \vec r
\frac{1}{H_A - E - \omega} \, \vec r \right| \phi_i \right> \,.
\end{multline}
We have allowed for a sign ambiguity on the right-hand side;
both signs would lead to the same Rabi frequency,
which is proportional to the absolute modulus of the 
transition matrix element.

In order to investigate whether the identity~\eqref{twogamma1}
holds, we specialize our general ``master identity'' 
given in Eq.~\eqref{genid}
to the case $| \phi_f \rangle \neq | \phi_i \rangle$,
\begin{align}
& \; \frac{\hbar^2}{m^2} \left< \phi_f \left| \vec p  \,
\frac{1}{H_A - E_n - \omega} \, \vec p \right| \phi_i \right> \,,
\nonumber\\[0.1133ex]
& \; = (E_f - E_i - \hbar \omega) (-\hbar \omega) 
\left< \phi_f \left| \vec r \, 
\frac{1}{H_A - E_n - \omega} \, \vec r \right| \phi_i \right> 
\nonumber\\[0.1133ex]
& \; \qquad + \left(\hbar \omega - \tfrac12 \, ( E_f - E_i) \right) \,
\left< \phi_f \left| \vec r^{\,2} \right| \phi_i \right> \,.
\end{align}
At exact resonance, i.e., for 
\begin{equation}
\hbar \omega = \hbar \omega_R = \tfrac12 \, ( E_f - E_i) \,,
\end{equation}
one has indeed
\begin{multline}
\label{twogammaR}
\frac{1}{m^2} \left< \phi_f \left| \vec p  \,
\frac{1}{H_A - E_n - \omega_R} \,
\vec p \right| \phi_i \right> \,,
\\[0.1133ex]
= -\omega_R^2 \left< \phi_f \left| \vec r \,
\frac{1}{H_A - E_n - \omega_R} \, \vec r \right| \phi_i \right> \,,
\end{multline}
which is exactly of the required form given in Eq.~\eqref{twogamma1},
for the case $\omega = \omega_R$.
Again, the minus sign does not influence 
the calculation of the Rabi frequency and is physically
irrelevant.

However, for $\omega \neq \omega_R$, i.e.,
off resonance, the two-photon transition rate as calculated
in the length gauge differs from the corresponding result
in the velocity gauge.
The identity~\eqref{twogamma1} does not hold
for $\omega \neq \omega_R$.
As already discussed in Sec.~\ref{sec2},
the ``gauge-noninvariance'' of {\em one}-photon transitions
is well known [see the discussion 
surrounding Eqs.~\eqref{fix1}---\eqref{fix2}].
For two-photon transitions,
the role of the inclusion of the entire spectrum of 
virtual atomic states has been somewhat 
unclear (see Refs.~\cite{PoTh1975,PoTh1978,CrTh1984}).
The initial conjecture
regarding the equality of the length and velocity-gauge
expressions can be traced to the paper by Geltman~\cite{Ge1963},
which treats a manifestly resonant process,
namely, the two-photon absorption and ionization of 
a ground-state hydrogen atom.
Geltman writes 
an expression which corresponds to the seagull term 
in Eq.~(4) of his paper, adds it to the length-gauge 
expression which he gives in his Eq.~(1), and 
asserts that the 
result is equal to the velocity-gauge result 
given in Eqs.~(3) and~(5) of Ref.~\cite{Ge1963}.
From the presentation, it is clear that 
Geltman's argument applies to a resonant process,
namely, the ionization rate of an initially ground-state 
atom by the absorption of two photons of frequency 
$\nu_2$, into a
continuum state of energy $E_f = E_i + 2 h \nu_2 > 0$
(in the notation of Ref.~\cite{Ge1963}).

\begin{figure}[t!]
\centerline{\includegraphics[width=0.9\linewidth]{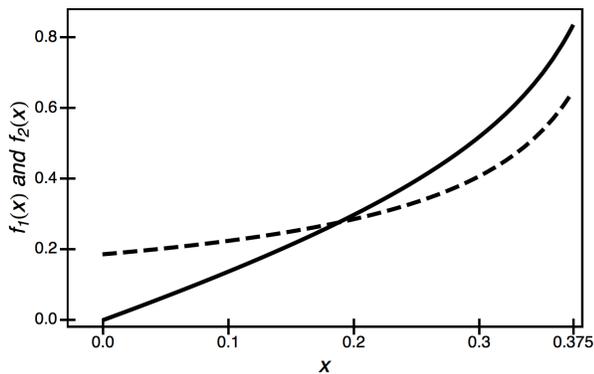}}
\caption{\label{fig1}
Comparison of the functions $f_1(x)$ 
(dashed line) and $f_2(x)$
(solid line) in the range 
$0 < x = \hbar \omega/(\alpha^2 m c^2) < 3/8$. 
The overlap occurs at 
$x = \tfrac{3}{16}$, which is the two-photon resonance condition.}
\end{figure}

\begin{figure}[t!]
\centerline{\includegraphics[width=0.9\linewidth]{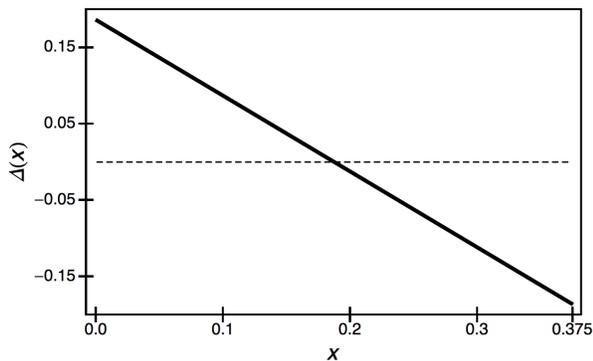}}
\caption{\label{fig2}
Gauge difference $\Delta(x) = f_1(x) - f_2(x)$. 
The zero occurs at $x = \tfrac{3}{16}$.}
\end{figure}

In view of advances in handling the 
Schr\"{o}dinger--Coulomb propagator~\cite{SwDr1991a,SwDr1991b,SwDr1991c},
and the calculation of 
energy-dependent matrix elements of the 
nonrelativistic propagator, which are necessary for analytic Lamb shift
calculations~\cite{Pa1993,JePa1996,JeSoMo1997}, it is feasible
to write analytic expressions 
for the two-photon transition matrix element
in the velocity and length gauges. 
We may define the dimensionless matrix element
\begin{align}
\label{matQ}
\calQ_{2S;1S}(\omega) =& \; 
\frac{(\alpha m c)^4}{3 m \hbar^2} \, Q_{2S;1S}(\omega) 
\\[0.1133ex]
=& \; \frac{(\alpha m c)^4}{3 m \hbar^2} \, 
\left< 2S \left| \vec r \,
\frac{1}{H_A - E_n - \omega} \,
\, \vec r  \right| 1S \right>  \,,
\nonumber
\end{align}
for which we may write the expression
\begin{multline}
\label{resQ}
\calQ_{2S;1S}(\omega) =
\frac{512 \, \sqrt{2} \, t^2}{729 \, (t-2)^3 \,(t^2 - 1)^2 \,(t+2)^2} \,
\\[2ex]
\times \left( 419 t^7 + 134 t^6 - 15 t^5 + 30 t^4 \right. \\
\left. + 60 t^3 - 120 t^2 - 32 t + 64 \right)
\\[2ex]
- \frac{4096 \, \sqrt{2} \;\;
{}_2 F_1\left(1, -t, 1-t, \frac{(1-t) \, (2-t)}{(1+t) \, (2+t)} \right)
}{3 \, (t^2-2)^3 \, (t^2-1)^2 } \,
\end{multline}
where
\begin{equation}
t = \sqrt{ 1 - \frac{2 \, \hbar \omega}{\alpha^2 m c^2} } \,.
\end{equation}
(All formulas pertain to atomic hydrogen, where we 
set the nuclear charge number equal to $Z=1$).
The dimensionless matrix element in the velocity gauge is
\begin{align}
\label{matP}
\calP_{2S;1S}(\omega) =& \;
\frac{m}{3 \hbar^2} \, P_{2S;1S}(\omega) 
\\[0.1133ex]
=& \; \frac{1}{3 m} \,
\left< 2S \left| \vec p  \,
\frac{1}{H_A - E_n - \omega} \,
\vec p \right| 1S \right>  \,,
\nonumber
\end{align}
for which the expression reads
\begin{multline}
\label{resP}
\calP_{2S;1S}(\omega) =
\frac{64\, \sqrt{2}}{81}  \,
\frac{t^2 (23 \, t^3 + 8 \, t^2 + t - 2)}{(t-2)^2\,(t^2-1) \,(t+2)}
\\[2ex]
- \frac{256 \, \sqrt{2} \;\;
{}_2 F_1\left(1, -t, 1-t, \frac{(1-t) \, (2-t)}{(1+t) \, (2+t)} \right)}%
{3 \, (t-2)^2 \, (t^2-1) \, (t+2)^2} \, .
\end{multline}
In Fig.~\ref{fig1}, we compare the expressions 
\begin{subequations}
\begin{align}
f_1 =& \; 
\calP_{2S;1S}(\omega) \,,
\\[2ex]
f_2 =& \; 
\frac{(E_{2S} - E_{1S} -\hbar\omega) \, (-\hbar\omega)}{(\alpha^2 \, m \, c^2)^2} \, 
\calQ_{2S;1S}(\omega)
\end{align}
\end{subequations}
in the range $\omega \in (0, E_{2S} - E_{1S})$,
as a function of $x = \hbar\omega/(\alpha^2 m c^2)$. 
The difference 
of the results given in Eqs.~\eqref{resP} and~\eqref{resQ},
\begin{align}
\Delta =& \; f_1 - f_2
\nonumber\\[0.1133ex]
=& \; \frac{m}{\hbar^2} 
\left( \hbar\omega - \tfrac12 (E_{2S}-E_{1S}) \right) 
\left< 2S \left| \vec r^{\,2} \right| 1S \right>
\nonumber\\[0.1133ex]
=& \; -\frac{512\, \sqrt{2}}{729} \left(x - \frac{3}{16}\right) \,,
\end{align}
is plotted in Fig.~\ref{fig2}.
At exact resonance, one has 
\begin{equation}
x = x_R = \frac{3}{16} \,,
\qquad
\Delta = 0 \,,
\end{equation}
as well as 
\begin{multline}
\calQ_{2S;1S}(\omega_R) = 
-\left(\frac{16}{3} \right)^2 \calQ_{2S;1S}(\omega_R) \\
= -\frac{2^{15}}{3^6} \,
\left[ 19 \, \sqrt{2} + 16 \, \sqrt{5}
+ 64 \, \sqrt{2} \;
\right. \\
\left.
\times \phi\left(-19 + 6 \, \sqrt{10}, 1, -2 \, \sqrt{\frac25}\right)
\right]
= -7.853\,655\,422 \,.
\end{multline}
Here, 
\begin{equation}
\phi(z,s,a) = \sum_{k=0}^\infty \frac{z^k}{(k + a)^s}
\end{equation}
is the Lerch $\phi$ transcendent.

With Lamb~\cite{La1952} and Kobe~\cite{Ko1978prl}, we note that the 
electric field is a gauge-independent quantity,
use the length-gauge expression and 
supply the prefactors in SI units in order 
to write the following expression for the 
Rabi frequency~\cite{HaEtAl2006},
\begin{align} 
\Omega =& \; 2 (2 \pi \beta_{2S;1S}) \, I_L \,,
\\[0.1133ex]
\qquad
\beta_{2S;1S}(\omega) = & \;
- \frac{e^2 \hbar}{\alpha^4 m^3 c^5 (4 \pi \epsilon_0)} \,
\calQ_{2S;1S}(\omega) \,.
\end{align}
An expansion of the Rabi frequency about 
resonance leads to the result
\begin{align}
\beta_{2S;1S}(\omega) =& \;
\beta_{2S;1S}(\omega_R) +
\left. \frac{\partial \beta_{2S;1S}(\omega)}{\partial \omega} 
\right|_{\omega = \omega_R} (\omega - \omega_R) 
\\[0.133ex]
=& \; \left[ 3.68111 \times 10^{-5} + 2.32293 \times 10^{-4} (x - x_R) \right] \,
\nonumber\\[0.133ex]
& \; \times \frac{{\rm Hz} \, {\rm m}^2}{{\rm W}^2} \,,
\end{align}
where $x = \omega/(\alpha^2 m)$ and $x_R = 3/16$.

A remark on two-color absorption is in order.
If an atom is simultaneously subjected 
to two laser fields of different frequencies
$\omega_1$ and $\omega$, which fulfill the resonance
condition $\omega_1 + \omega_2 = E_f - E_i$, 
then gauge invariance is restored. 
On the basis of Eq.~\eqref{genid},
this is verified (again for two-photon resonance) as follows,
\begin{multline}
P(\omega_1) + P(\omega_2) 
= (E_f - E_i - \hbar\omega_1) (-\hbar\omega_1) \, Q(\omega_1)
\\
(E_f - E_i - \hbar\omega_2) (-\hbar\omega_2) \, Q(\omega_2)
\\
+ \left[\hbar\omega_1 - \tfrac12 ( E_f - E_i) \right] \,
\left< \phi_f \left| \vec r^{\,2} \right| \phi_i \right>
\\
+ \left[\hbar\omega_2 - \tfrac12 ( E_f - E_i) \right] \,
\left< \phi_f \left| \vec r^{\,2} \right| \phi_i \right>
\\
= - (\hbar\omega_1) (\hbar\omega_2) \, [ Q(\omega_1) + Q(\omega_2) ] \,.
\end{multline}
The relation $P(\omega_1) + P(\omega_2) = 
- (\hbar\omega_1) (\hbar\omega_2) \, [ Q(\omega_1) + Q(\omega_2) ]$
is equivalent to the gauge invariance of the 
resonant two-color, two-photon transition.
In Tables~I and II of Ref.~\cite{BaFoQu1977}, the 
authors present resonant two-color, two-photon matrix 
elements which in our notation would read
\begin{equation}
\label{twoterms}
\calQ_{\rm 2c;2\gamma}(\omega_1) =
\frac34 \, \left[ \calQ_{2S;1S}(\omega_1) + \calQ_{2S;1S}(\omega_2) \right] \,,
\end{equation}
where, again, $\omega_2 = E_f - E_i - \omega_1$.
For example, the gauge-invariant resonant two-color 
result at frequency $\frac{28}{15} \omega_R$ reads as 
\begin{multline}
\calQ_{2c;2\gamma}\left(\omega_1 = \frac{28}{15} \omega_R = 
\frac{7}{20} \frac{\alpha^2 m c^2}{\hbar}\right) 
= -\frac{160\,000}{343} 
\\
\times \Biggl[ 157 \sqrt{2} + 56 \sqrt{15} + 2 \sqrt{190} 
\\
+ 560 \, \sqrt{2} \; 
\phi\left( \tfrac17 \, (-263 + 48 \, \sqrt{30}),
1, -\sqrt{\frac{10}{3}} \right) 
\\
+ 64 \, \sqrt{2} \;
\phi\left( \tfrac17 \, (-848 + 87 \, \sqrt{95}),
1, -2 \sqrt{\frac{5}{19}} \right) \Biggr]
\\
= -62.659\,473\,633 \,,
\end{multline}
verifying the fifth entry in the last row
of Tables~I and II of Ref.~\cite{BaFoQu1977}.
Diagrammatically, the two terms in Eq.~\eqref{twoterms}
correspond to photon absorption processes 
with two different possible time orderings
of the absorptions of photons with 
frequencies $\omega_1$ and $\omega_2$.

%
%
\section{Conclusions}

In the current paper,
we (re-)examine the transformation 
from the length to the velocity gauge in Sec.~\ref{sec2},
and recall that the 
length-gauge and velocity-gauge 
Hamiltonians are not related by a 
unitary transformation.
Furthermore, we show that the 
physical interpretation of a quantum mechanical 
operator depends on the gauge, 
vindicating arguments given by Lamb~\cite{La1952} and 
Kobe~\cite{Ko1978prl} regarding 
the applicability of the length gauge off resonance.
In Sec.~\ref{sec3},
we consider the ac Stark shift as a paradigmatic 
example of a physical process invariant under 
an ``extended'' gauge transformation.
Specifically, in atomic hydrogen,
we rederive the known result~\cite{HaJeKe2006,HaEtAl2006}  that 
the ac Stark shift formulated in the length gauge 
is equal to the velocity-gauge
expression, even if the gauge transformation of the 
wave function is ignored.
The derivation is based on the ``master identity'' given in
Eq.~\eqref{genid}.
In Sec.~\ref{sec4}, we investigate
the two-photon transition matrix element,
where the ``extended gauge invariance'' does not hold
off resonance.
The derivation again profits from the general identity~\eqref{genid},
which can be applied to both 
of the problems studied in Secs.~\ref{sec3} and~\ref{sec4};
its validity is
verified on the basis of analytic and numerical calculations
[see Eqs.~\eqref{resQ},~\eqref{resP} as well as 
Figs.~\ref{fig1} and~\ref{fig2}].
In retrospect, it would have seemed somewhat surprising
if extended gauge invariance had been applicable
to two-photon transitions (under the inclusion of 
all possible virtual, intermediate states) 
but failed for one-photon transitions
[see Eqs.~\eqref{fix1}---\eqref{fix2}].
We conclude that for two-photon transitions, the
length gauge needs to be used off resonance,
just as for one-photon absorption.

Yet, for two-color, two-photon absorption with 
the sum of the two photon frequencies adding up to the 
exact resonance frequency,
extended gauge invariance again holds (see Sec.~\ref{sec4}
and Ref.~\cite{BaFoQu1977}).

A few explanatory remarks are in order.
We have seen that extended gauge invariance is restored at exact 
resonance, for both one- as well as (one-color and two-color) 
two-photon transitions.
Mathematically, extended gauge invariance is restored at resonance
in view of commutation relations, notably,
$\vec p = \frac{\ii m}{\hbar} [H_A, \vec r]$,
where $H_A$ is the atomic Schr\"{o}dinger--Coulomb Hamiltonian.
Physically, extended gauge invariance holds because 
processes at exact resonance, or, processes which 
involve energy shifts, can be formulated 
using a form of the interaction where the 
fields and potentials are adiabatically
switched off in the infinite future and in the infinite past,
using a damping term of the form $\exp(- \epsilon | t |)$.
The gauge transformation of the wave functions 
(in and out states) then proceeds in the 
distant past and future, where the fields are switched
off and the gauge transformation is just the identity.
For one- and two-photon transitions, the necessity to 
introduce the damping terms is inherent 
to the formulation of Fermi's Golden Rule,
which describes transition rates at exact resonance,
where the initial and final states fulfill an 
energy conservation condition [see Refs.~\cite{Sa1994Mod,Sa1967Adv}.

The ac Stark shift can be formulated 
using the Gell--Mann--Low theorem
[see Eqs.~(19) and~(21) of Ref.~\cite{HaJeKe2006}], 
in which case one uses a time
evolution operator that evolves the wave function 
from the infinite past to the present,
with the interactions being switched off for $t \to -\infty$.
Within the Gell--Mann--Low formalism, 
the gauge transformation of the wave function in the infinite past 
amounts to the identity transformation,
because the interactions are adiabatically
switched off in this limit. The 
extended gauge invariance of those physical processes whose
description allows such an adiabatic damping,
thus finds a natural explanation.
For one- and two-photon transitions off resonance,
however, the quantum dynamics are instantaneous,
and the physical interpretation of the 
operators must be carefully restored. 
In this case, only the length gauge 
provides a consistent physical description (see the 
discussion in Sec.~\ref{sec2}).

One might thus ask if the velocity gauge has any advantages 
in the physical description of laser-related processes.
The answer can be given as follows.
There are $S$-matrix elements in the so-called 
strong-field approximation
whose evaluation becomes easier in the velocity gauge.
In this case, the in- and out-states are 
asymptotic states [the $S$ matrix is a time evolution
operator from the infinite past to the infinite future].
Indeed, as stressed by Reiss in Eqs.~(29) and~(31) 
of Ref.~\cite{Re2013}, the Volkov state in a strong laser field
is much easier to formulate in the velocity gauge,
and consequently, $S$-matrix calculations should 
preferentially be 
done in this gauge [see also Refs.~\cite{Re1992,CoLa1996,Re2013}].
In the formulation of the $S$ matrix,
one canonically uses infinitesimal damping parameters
[see Ref.~\cite{MoPlSo1998}],
and thus, extended gauge invariance is restored.
We conclude that 
the choice of gauge in these cases should be made according 
to practical considerations,
and in strong laser fields, 
the velocity gauge provides for the 
most simple computational framework.

\section*{Acknowledgments}

This research has been supported by the National 
Science Foundation (grant PHY--1403973).

\appendix

\end{document}